\documentclass[a4paper,11pt]{article}
\pdfoutput=1 

\usepackage{jinstpub} 

\usepackage{booktabs}
\usepackage{xspace}
\usepackage{siunitx}
\usepackage{lineno}

\title{\boldmath EUDAQ2 -- A Flexible Data Acquisition Software Framework for Common Test Beams}

\author[a,1]{Y.~Liu\note{Corresponding author.}}
\author[b]{M.~S.~Amjad}
\author[c]{P.~Baesso}
\author[c]{D.~Cussans}
\author[a]{J.~Dreyling-Eschweiler}
\author[a]{R.~Ete}
\author[a]{I.~Gregor}
\author[a]{L.~Huth}
\author[d]{A.~Irles}
\author[a]{H.~Jansen}
\author[a]{K.~Krueger}
\author[e,a]{J.~Kvasnicka}
\author[a,f]{R.~Peschke}
\author[a]{E.~Rossi}
\author[g]{A.~Rummler}
\author[a]{F.~Sefkow}
\author[a]{M.~Stanitzki}
\author[b,a]{M.~Wing}
\author[a]{M.~Wu}

\affiliation[a]{Deutsches Elektronen-Synchrotron,\\Notkestr. 85, 22607 Hamburg, Germany}
\affiliation[b]{Department of Physics and Astronomy, University College London,\\ Gower Street, London WC1E 6BT, United Kingdom}
\affiliation[c]{University of Bristol,\\Tyndall Avenue, Bristol BS8 1TL,United Kingdom}
\affiliation[d]{Laboratoire de l'Acc\'elerateur Lin\'eaire (LAL), CNRS/IN2P3 et Universit\'e de Paris-Sud XI,\\Centre Scientifique d'Orsay, B\^atiment 200, BP 34, F-91898 Orsay, CEDEX, France}
\affiliation[e]{Institute of Physics of the Czech Academy of Sciences, \\Na Slovance 2, 18221 Prague 8, Czech Republic }
\affiliation[f]{Department of Physics and Astronomy, University of Hawai'i at M\={a}noa, \\Watanabe 416, 2505 Correa Road, Honolulu, HI 96822, USA}
\affiliation[g]{European Organization for Nuclear Research (CERN),\\1211 Geneva 23, Switzerland}

\emailAdd{yi.liu@desy.de}

\newcommand{\eudaq}{\textsc{EUDAQ}\xspace}
\newcommand{\eudaqii}{\textsc{EUDAQ2}\xspace}

\abstract{
The data acquisition software framework, \eudaq, was originally developed to read out
data from the EUDET-type pixel telescopes.  This was successfully used in many
test beam campaigns in which an external position and time reference were required.  The
software has recently undergone a significant upgrade, \eudaqii, which is a generic, modern and 
modular system for use by many different detector types, ranging from tracking detectors to calorimeters.
\eudaqii is suited as an overarching software that links individual detector readout systems and simplifies 
the integration of multiple detectors. The framework itself supports several triggering and event building modes.
This flexibility makes test beams with multiple detectors significantly easier and more efficient,
as \eudaqii can adapt to the characteristics of each detector prototype during testing.
The system has been thoroughly tested during multiple test beams involving different detector prototypes. \eudaqii has now been 
released and is freely available under an open-source license.
}

\newcommand{\desyii}{{\mbox{DESY II}}\xspace}
\newcommand{\diitbf}{{\desyii Test Beam Facility}\xspace}

\newcommand{\EUDAQ}{\textsc{EUDAQ}\xspace}
\newcommand{\EUDAQII}{\textsc{EUDAQ2}\xspace}
\newcommand{\EUTELESCOPE}{\textsc{EUTelescope}\xspace}
\newcommand{\python}{\textsc{Python}\xspace}

\newcommand{\ROOT}{\textsc{ROOT}\xspace}
\newcommand{\LCIO}{\textsc{LCIO}\xspace}
\newcommand{\CMAKE}{\textsc{CMake}\xspace}
\newcommand{\epem}{\mbox{${\rm e}^+{\rm e}^-$}}

\newcommand{\EUDET}{\textsc{EUDET}\xspace}
\newcommand{\AIDA}{\textsc{AIDA}\xspace}

\newcommand{\RunControl}{\texttt{Run\,Control}\xspace}
\newcommand{\Producer}{\texttt{Producer}\xspace}
\newcommand{\Producers}{\texttt{Producers}\xspace}
\newcommand{\DataCollector}{\texttt{Data\,Collector}\xspace}
\newcommand{\DataCollectors}{\texttt{Data\,Collectors}\xspace}

\newcommand{\LogCollector}{\texttt{LogCollector}\xspace}
\newcommand{\OnlineMonitor}{\texttt{OnlineMonitor}\xspace}
\newcommand{\Monitor}{\texttt{Monitor}\xspace}

\newcommand{\RAWDATAEVENT}{\textit{RawEvent}\xspace}
\newcommand{\STDDATAEVENT}{\textit{StdEvent}\xspace}
\newcommand{\RAWDATABLOCK}{\textit{RawDataBlock}\xspace}
\newcommand{\DATACONVERTER}{\texttt{DataConverter}\xspace}
\newcommand{\EVENT}{\textit{Event}\xspace}
\newcommand{\STATUS}{\textit{Status}\xspace}
\newcommand{\SUBEVENT}{\textit{SubEvent}\xspace}

\newcommand{\SERIALIZABLE}{\textit{Serializable}\xspace}

\newcommand{\INIT}{\textit{INIT}\xspace}
\newcommand{\CONFIG}{\textit{CONFIG}\xspace}
\newcommand{\START}{\textit{START}\xspace}
\newcommand{\STOP}{\textit{STOP}\xspace}
\newcommand{\RESET}{\textit{RESET}\xspace}
\newcommand{\TERM}{\textit{TERM}\xspace}

\newcommand{\UNINIT}{\textit{UNINIT}\xspace}
\newcommand{\UNCONF}{\textit{UNCONF}\xspace}
\newcommand{\CONF}{\textit{CONF}\xspace}

\newcommand{\RUNNING}{\textit{RUNNING}\xspace}
\newcommand{\ERROR}{\textit{ERROR}\xspace}

\newcommand{\uinteightt}{\texttt{uint8t}\xspace}
\newcommand{\KPIX}{{KPiX}\xspace}
\newcommand{\LYCORIS}{\textsc{Lycoris}\xspace}
\newcommand{\MIMOSA}{\textsc{Mimosa}\xspace}
\newcommand{\MIMOSAXXVI}{\textsc{Mimosa26}\xspace}


\arxivnumber{1907.10600}

\begin{document}
\maketitle
\flushbottom

\section{Introduction}
\label{sec:intro}
The next generation of particle physics experiments requires detectors with an outstanding
performance to be designed, built and tested. The challenges involve spatial resolutions to the micron level, 
picosecond timing resolution and more on-detector intelligence. At the same time, the material budget needs to be further reduced compared 
to present systems, which requires novel solutions for the readout, powering and cooling of the detectors.
As a part of any successful R\&D program a set of test beams for each new detector are required to demonstrate its capabilities and performance.

In order to facilitate these detector test beams, high resolution pixel beam telescopes, 
the so-called \EUDET-type pixel beam telescopes~\cite{jansen2016} (Sect.~\ref{sec:telescopes}) have been developed 
as a common infrastructure available to any R\&D group.
The accompanying data acquisition software, \EUDAQ~\cite{eudaq12019}, was originally developed to read out the 
data from the \EUDET-type pixel telescopes and was therefore closely connected to its data acquisition~(DAQ) architecture. During a 
decade of usage many user groups have integrated their DAQ system into \EUDAQ and have successfully combined 
their data with the telescope data, based on common trigger IDs. Together with the \EUTELESCOPE software package 
, a common pixel telescope data analysis framework, the \EUDET-type pixel beam telescopes offer the whole infrastructure 
for detector development from the initial measurements to the final results.
Furthermore the availability of these telescopes at the test beam facilities at CERN, DESY, ELSA(Bonn) and SLAC 
provided users the possibility to move their setup between test beams and use the same interface to the telescopes.

The recent development of \EUDAQII  provides a more flexible DAQ
software framework for operating detector prototypes at test beams worldwide. The 
emerging need for running several detectors together with a telescope as a so-called 
system test is extending the use case way beyond the original conception of 
\EUDAQ. Being a major upgrade to \EUDAQ, \EUDAQII has been completely rewritten using modern C++ 
and was designed to become even more versatile and usable for an extended range of 
detectors. The modular and cross-platform data acquisition framework serves as a 
flexible and simple-to-use data taking software for the \EUDET-type pixel beam 
telescopes while allowing for the easy integration of many other detectors. 
\EUDAQII is freely available~\cite{eudaq_github} and distributed under the LGPLv3~\cite{LGPLv3} open-source license.

\section{\EUDAQII Architecture}

\EUDAQII, like its predecessor \EUDAQ, is implemented in C++, taking advantage of many of the powerful features provided by 
the C++11 standard~\cite{ISO:2012:III}. 
Compatibility across operating systems and compilers is one of the \EUDAQII design principles, 
hence it only uses standard C++ language features and POSIX~\cite{8277153} system routines. 
Therefore, \EUDAQII can run natively on Linux, MacOS and Windows. 
To build an \EUDAQII system, different compilers such as GCC, LLVM/Clang and MS Visual~C++ are supported. 
The build and installation processes are configured using \CMAKE~\cite{cmake}.
\EUDAQII is a distributed DAQ framework with its communication protocol running on a custom TCP/IP stack.

\begin{figure}[htb]
  \begin{center}
    \includegraphics[width=0.9\textwidth]{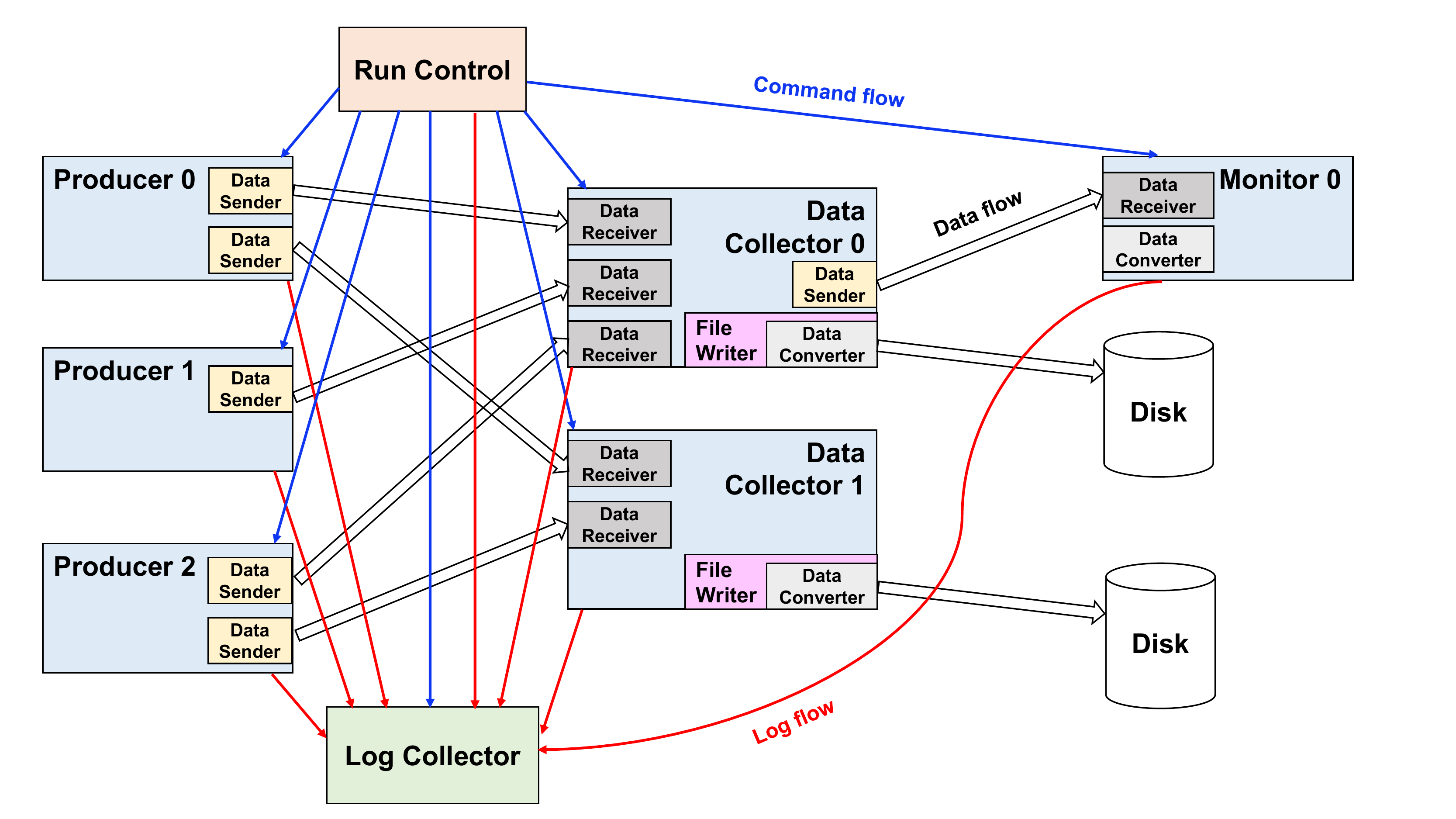}
    \caption{A schematic view of the \EUDAQII architecture.}
    \label{fig:arch:eudaq}
  \end{center}
\end{figure}

Figure \ref{fig:arch:eudaq} shows a schematic overview on the \EUDAQII framework: 
Each component of \EUDAQII can run anywhere on the network on 
separate machines, entirely operating-system independent, and connect to each other using the 
\EUDAQII data taking setup at run-time. Within the \EUDAQII framework, each detector 
hardware is being controlled and read out by an individual \EUDAQII instance.
 At the same time, the data streams from different detectors can be merged using 
well-defined synchronization mechanisms and stored to disk.

\subsection{Distributed Run-time Roles}
A typical \EUDAQII setup is split into several run-time instances with different roles, each communicating via the network.
Table \ref{tab:arch:roles} gives an overview of the roles of each \EUDAQII instance.

\begin{table}[htb]
\caption{\EUDAQII Roles.}
\begin{center}
\begin{tabular}{lp{6cm}}
\EUDAQII Role & Description\\
\toprule
\RunControl & Central controller for full \EUDAQII system\\
\Producer & Controls an individual detector, and sends detector data to \EUDAQII \\
\DataCollector & Collects and merges data from individual \Producers, then stores the data to disk file \\
\LogCollector & Collects log messages from the entire \EUDAQII system\\
\Monitor & Online monitoring of  data quality\\
\texttt{Offline Tools} & Fast offline data conversion and data analysis \\
\bottomrule
\end{tabular}
\label{tab:arch:roles}
\end{center}
\end{table}

The \RunControl is at the core of each \EUDAQII system.
Every \EUDAQII system requires exactly one running instance of the \RunControl.
All other \EUDAQII participating instances must be made aware of the network location of the \RunControl and 
announce themselves to the \RunControl at startup.
The \RunControl reads and parses the initialization/configuration files and distributes the corresponding settings
to each connected \EUDAQII instance according to the instance's run-time name.
It also serves as user interface via an optional graphical interface.
The \RunControl is responsible for the start and stop the data taking.

A \Producer controls underlying detector hardware which participates in the \EUDAQII data taking.
Each detector hardware needs to have a dedicated \Producer. Normally, there are several \Producers running in parallel,
as the data taking during a test beam consists of several detectors.
The \Producers feed  the detector data into the distributed \EUDAQII framework.
A \Producer also responds to \RunControl commands and manages the detector hardware accordingly.
The \Producer is the only part where the user is required to make a dedicated and hardware-specific implementation, 
technically by employing the C++ polymorphism mechanism.
A \Producer may support both an "internal" and and "external" loop for interacting with the hardware and retrieving data, when it is available. 
For the "internal" loop, the \Producer provides a loop/thread internally, manages the \START and \STOP commands and 
handles the error exceptions. This is the simplest way for users to integrate their hardware into \EUDAQII. 
This might not be suitable for certain DAQ systems, which provide their own read-out loop and state-machine. 
Therefore another mode, the "external" loop mode, where \EUDAQII is not managing the readout but merely receives data, 
whenever it is made available, is provided.

The \DataCollectors collects the data from the individual detectors via their \Producers and write data to disk.
A \Producer can be configured to send data to one or several \DataCollector instances, depending on the user needs during data taking.
A \DataCollector can be configured to receive data from one or several \Producers, hence supporting very flexible 
ways of event-building when using different detectors.
The synchronization of the data from different \Producers can be done either using
straight-to-disk, skipping event validation and synchronization, sync-with-arrival-order, sync-with-trigger and sync-with-timestamp,  
all of them fully supported by \EUDAQII.
Basic sanity checks such as testing the consistency of event numbers or TriggerIDs are implemented by default.
The API of the \DataCollector allows straightforward inclusion of additional synchronization methods if required by a user, see section~\ref{sec:extension}.

The \LogCollector gathers logging information from all \EUDAQII instances and displays them centrally in one unified logging window. 
A single log file is stored along with the corresponding data file for later reference.
This simplifies the tracking of problems encountered during data-taking.
Only a single instance of the \LogCollector is allowed in any \EUDAQII setup.
If a setup does not provide a \LogCollector instance during the run, the log messages just go to the local screen where an \EUDAQII instance runs.

The \Monitor analyses the incoming data stream and generates a set of histograms, ensuring  data quality online. 
Hardware failures of setup issues can be tracked down efficiently utilizing the \Monitor. 
A legacy \OnlineMonitor from \EUDAQ is shipped to maintain compatibility and to simplify migration to \EUDAQII. 
The \Monitor is able to run offline with disk data as well, which allows for quick data quality verification. 

Data decoding can be performed either online or offline. Independently, the \DATACONVERTER is the only point
where specific decoding routines need to be implemented. The \DATACONVERTER to be called for a specific 
event is derived from the event type. A corresponding \DATACONVERTER can be retrieved and called by 
the \Monitor and other offline analysis tools.

\subsection{State \& Command Model}

\begin{figure}[htbp]
	\centering
	\includegraphics[width=\textwidth]{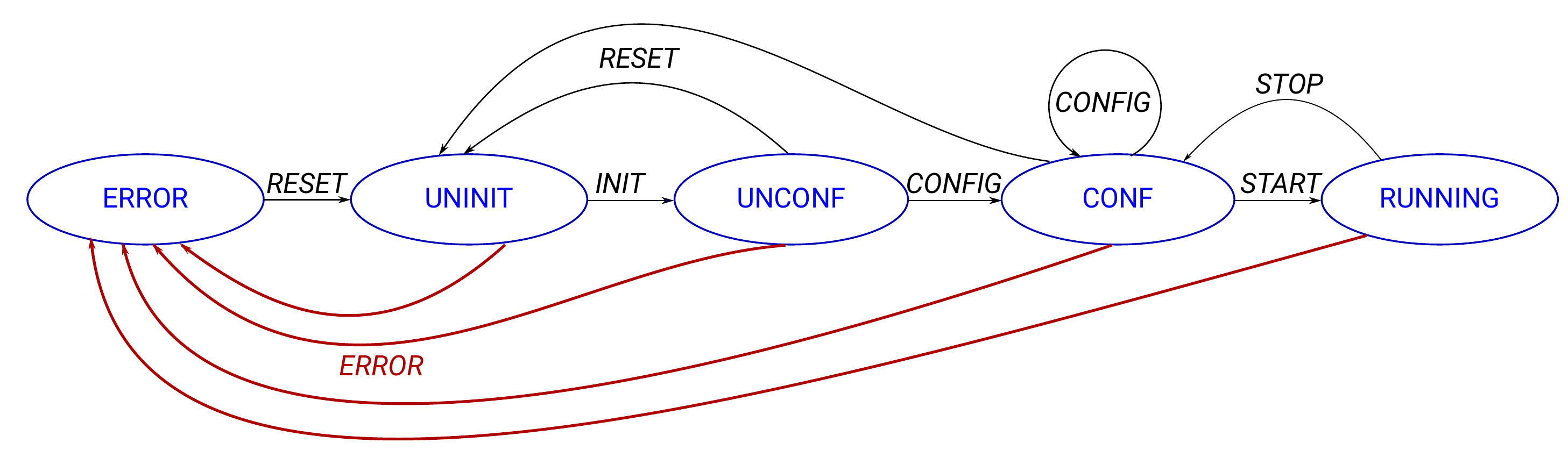}
	\caption{Finite State Machine of \EUDAQII. States are displayed in blue, 
	\textit{commands} in black and error handling in red.  For the successful 
	initialization and configuration of  \EUDAQII, valid initialization and configuration files need to be provided. 
	\label{fig:arch:sm}}
\end{figure}

\begin{table}[htbp]
\caption{The \EUDAQII commands to trigger a change between \EUDAQII States \label{tab:arch:commands}.}
\begin{center}
\begin{tabular}{lllp{5cm}}
 Command &State         &State      &  Command Description \\
	 &before command&after command& \\
\toprule
\INIT 	& \UNINIT 		& \UNCONF	& Initialize using the initialization file\\
\CONFIG & \UNCONF/\CONF 	& \CONF 	& Configure using the configuration file\\
\START 	& \CONF 		& \RUNNING 	& Start up a new run \\
\STOP 	& \RUNNING 		& \CONF 	& Stop the current run \\
\RESET 	& \ERROR/\CONF/\UNCONF 	& \UNINIT	& Reset all running components\\
\TERM 	& all except \RUNNING 	& 	 	& Terminate \EUDAQII \\
\bottomrule
\end{tabular}
\end{center}
\end{table}
Any \EUDAQII instance has to maintain a run-time state. To change between the individual \EUDAQII states, a set of \EUDAQII commands is available, 
which is shown in Table~\ref{tab:arch:commands}. The finite state machine including allowed transitions is shown in Figure~\ref{fig:arch:sm}. 

Each \EUDAQII instance reports  the state \UNINIT to the \RunControl at startup.
An issued \INIT command by the \RunControl triggers the initialization of an instance, which changes the state to 
\UNCONF, if no errors occur.
A subsequent successful execution of the \CONFIG command leads to a reported \CONF state.
If all instances are in a \CONF state, the \RunControl is able to \START the readout, changing the \EUDAQII state to \RUNNING.
The \STOP command stops the data taking and the \EUDAQII instance keeps waiting for a new run as soon as another \START command arrives.
In case a new configuration file needs to be used and distributed to the \Producers,
a \CONFIG command needs to be executed between the \STOP and the next \START.
In case of an unexpected error, the state \ERROR is reported.
The only way to recover from an \ERROR state is to execute the \RESET command which can reset the system to \UNINIT state. 
Errors are handled by an exception mechanism.
The \TERM command terminates the complete system, closing all \EUDAQII instances.
It is worth noting that the \INIT command can only be executed once while the \CONFIG command can be issued several times.

It is the task of the \RunControl instance to issue the commands to alter the states of all participating instances and check/validate the return codes of all instances.
If an error occurs during the processing of a command by a \Producer, e.g. during the configuration of the underlying detector hardware, 
an exception can be issued, which is caught and automatically converted to an error state.
The error state is part of the return information collected by \RunControl. The \LogCollector receives a detailed error report automatically.
Commands pushed from \RunControl to each \Producer are processed in the sequence they were received.
a request to retrieve the current state from all 
the connected \EUDAQII instances is sent periodically and the state is updated accordingly, hence providing a heartbeat of the entire system.

The configuration of the \EUDAQII data acquisition framework is performed via two global configuration files.
One of them is used by the \INIT command, and the second for the \CONFIG command.
Both configuration files are stored as plain text and are divided into named sections for the individual run-time instance.
Each configuration section, named by combination of role and run time name of the running destination instance,
contains a list of parameter-value pairs which can be either mandatory or optional.

\subsection{Data Model}
Data objects are sent between various \EUDAQII instances in the same way as State and Commands. 
The data objects therefore need the capability to be serialized. When a data object is
serialized, all the crucial data of this data object is fed to a serialized memory section which then
can be sent as a plain binary data stream to another application and reconstructed as a copy of the original data object.
Technically, the \EUDAQII native data file format is a collection of serialized binary data streams from data objects

In \EUDAQII, the \SERIALIZABLE class is implemented by the \EVENT class. The \EVENT class stores data from a detector.
The basic \EUDAQII \EVENT implementation provides a few general variables as the \EVENT object 
is not required to be aware of all the details about each individual detector.
But it is mandatory for each \EVENT Object to have a few parameters set properly. 
Table~\ref{tab:arch:datamodell} lists all member variables of a \SERIALIZABLE \EVENT object and the members which \EUDAQII sets automatically.

\begin{table}
\caption{\EUDAQII Data Model\label{tab:arch:datamodell}.}
\begin{center}
\begin{tabular}{llc}
Variable &  Provided by User& Set by \EUDAQII automatically\\
\toprule
EventType & no & yes \\
RunNumber & no & yes \\
EventNumber & no & yes \\
Timestamp & optional & no \\
TriggerNumber & optional & no \\
\RAWDATABLOCK & yes & no \\
Tags & optional & no \\
SubEvents & optional & no \\
\bottomrule
\end{tabular}
\end{center}
\end{table}

The \RAWDATABLOCK contains the corresponding detector data stored in a detector-specific format, encoded as an array of \uinteightt. 
It is the responsibility of each user to provide the necessary data decoders using a \DATACONVERTER.
A pair of timestamps defines the start and end time of the \RAWDATABLOCK and a trigger number identifies the trigger sequence defined by the hardware setup. 
The timestamps and trigger number are per se optional, but are necessary to be set if they are to be used to synchronize data from multiple data streams. 
In this case, the raw measurement data has to be partly processed and decoded since the timestamps and trigger numbers are part of the detector raw data.
However, to minimize the CPU consumption due to data processing by the \Producer, full decoding of the detector raw data can be postponed to the 
online monitoring stage or the offline analysis.
It is possible to encapsulate several events inside an \EVENT object, the so-called \SUBEVENT objects.
This can be the case if the detector hardware controls multiple subsystems or has a cascaded data structure.
The \SUBEVENT objects are stored in the parent \EVENT object and are therefore also \SERIALIZABLE.

\subsection{Network communication}
The distributed communication is running using only TCP/IP with a custom protocol, based on the developments for \EUDAQ .
Replacing the custom protocol with a more modern, performance optimized network protocol, allowing for higher data 
throughput and restoring lost packages has been investigated, but not yet implemented.
This can be achieved with the existing abstraction layer and can be implemented 
as an extension library, without influencing any user code, see section~\ref{sec:extension}. 
We foresee this for one of the future \EUDAQII releases depending on user needs.

\subsection{The Graphical user interfaces}
\begin{figure}[htbp]
  \begin{center}
    \includegraphics[width=0.9\textwidth]{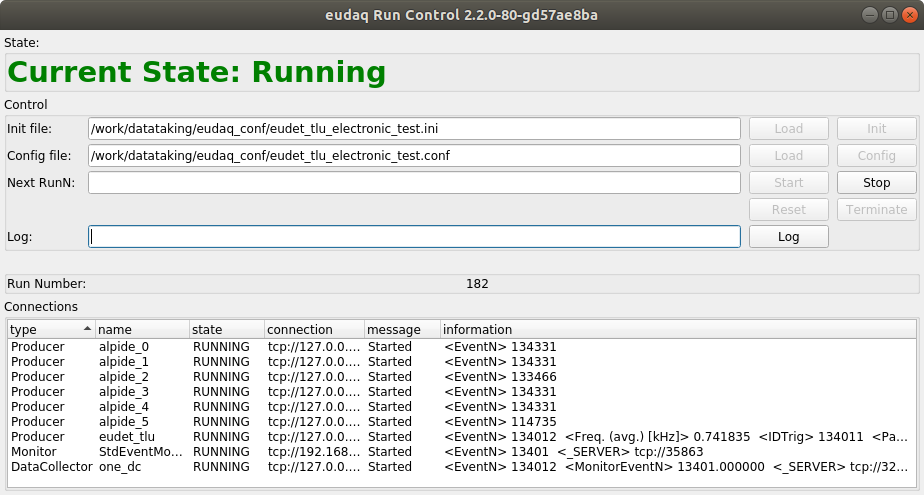}
    \caption{The Graphical user interface of the \EUDAQII \RunControl.}
    \label{fig:arch:eudaqruncontrol}
  \end{center}
\end{figure}

\EUDAQII provides Graphical user interfaces (GUIs) based on the Qt framework \cite{qt5}, which is a free 
and open-source widget toolkit to create GUIs for cross-platform applications. The Qt-based 
versions of the \EUDAQII \RunControl (see Fig.~\ref{fig:arch:eudaqruncontrol}), \LogCollector and 
a \Monitor are available as part of the \EUDAQII package together with a command line version.

\section{Integration with User code}
\begin{figure}[htbp]
  \begin{center}
    \includegraphics[width=0.7\textwidth]{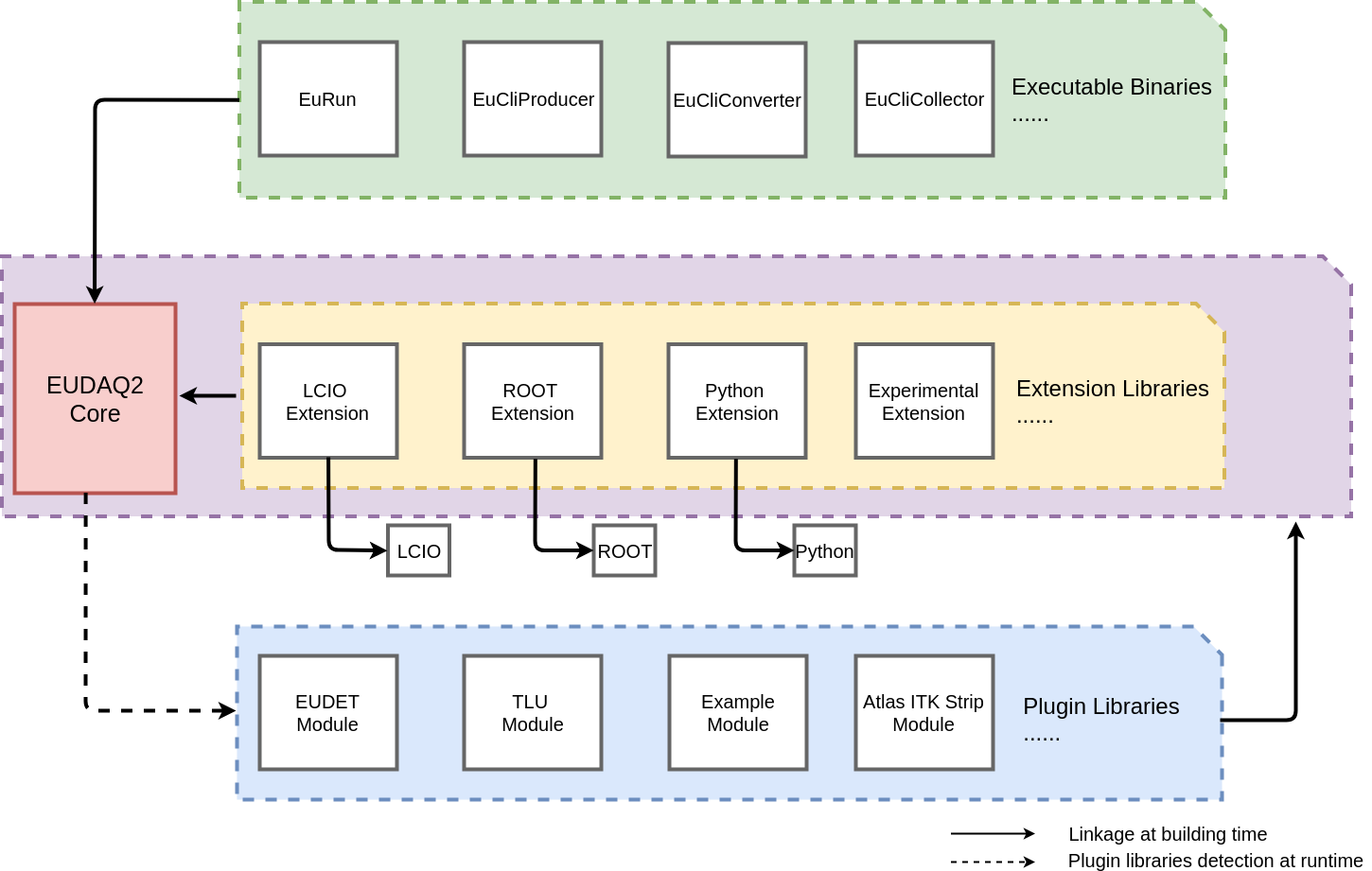}
    \caption{Layout of \EUDAQII, showing the relation between the \EUDAQII core libraries and the Executables, Extensions and Plugins.}
    \label{fig:libs}
  \end{center}
\end{figure}

To keep the pure C++ \EUDAQII core clearly separated from all user code, which may depend on non-standard libraries, the 
\EUDAQII binary library is split into a core library and and optional module/extension libraries.
Figure~\ref{fig:libs} shows this relation libraries with respect to the \EUDAQII core library.
A \CMAKE configure file to support user code integration against the \EUDAQII core is supplied.

\EUDAQII is not compatible with \EUDAQ, as the API has been changed and several interface methods have been deprecated.
However, the migration to \EUDAQII is straightforward and requires only minor changes in the user code, as the exception and error handling has changed.
To avoid accidental misuse of \EUDAQ user code in \EUDAQII, all interface functions have been renamed.

\subsection{Modular Plugins}
Modular plugins are used to interact with the user hardware, the DAQ systems, or to adjust the \EUDAQII core functionalities to meet certain user requirements.
A typical \EUDAQII modular plugin contains a \Producer, a \DataCollector and specific hardware code provided by the user. Optionally a modified 
\RunControl can also be defined as a module. External libraries or specific dependencies of a module have to be handled by the 
user by adjusting the module's \CMAKE file. A shared library is created for each modular plugin at compile time. 
This library is loaded at the startup of \EUDAQII run-time instances.

\EUDAQII takes advantage of the Object Oriented Programming (OOP)~\cite{doi:10.1080/03081079.2010.539975} capabilities of C++ to create 
objects like \Producers and \DataCollectors, which inherit from the base classes by method dispatching.
Using factories instead of constructors allows the usage of polymorphism for the object creation.

There are a few generic \EUDAQII executables, so-called portals, which provide a generic way
to create a user instance, e.g. a \Producer. Examples for such portals are the the {\tt euCliProducer} and {\tt euCliConverter} (see Fig.~\ref{fig:libs}), 
which are generic utilities. After a \EUDAQII portal executable has created an \EUDAQII instance, the portal executable only interacts with the 
\EUDAQII core library, while all user-specific functionality required by each instance is implemented in modular plugins.
The \EUDAQII core then takes care of searching and loading all available user plugin modules located in the specified directories.
The communication between the \EUDAQII core and its plugins is based on derived OOP classes.
Therefore, the \EUDAQII core can steer any user plugin binary without having knowledge of the hardware/plugin specific dependencies,
both during build and during runtime. This encapsulation enables simple changes in the prototypes and class hierarchies.

\subsection{Extension library}
\label{sec:extension}
Supporting \EUDAQII on multiple operation systems requires a compact core package 
without any external dependencies. However, this limits the available functionality, 
which can be an issue in certain use cases. To increase the flexibility, the 
core library is very modular with a minimal coupling between the major components. 
This allows for a simple replacement of individual components using extension libraries. An 
extension library provides additional optional core features, which might then
require external dependencies. These extensions provide a convenient way to 
extend the functionality of \EUDAQII, without changing the core libraries themselves.

The data for example can only be stored as a raw and uncompressed serial-event object stream within the standard \EUDAQII core. 
Extensions can be used to write out data using different forms: The \LCIO \cite{Gaede:2003ip,Aplin:2012kj} extension library 
for example allows for writing data as \LCIO-formatted data, which is widely used in the Linear Collider community. 
Developments of an extension to support data storage in \ROOT~{\tt TTrees}~\cite{Brun:1997pa,Antcheva:2009zz} is currently work in progress.

\subsection{\python interface}
\python has become a very popular programming language in the particle physics community.
Therefore, a \python wrapper based on {\tt pybind11}~\cite{pybind11_cite}, which is a lightweight, header-only library that exposes C++ types to \python, is provided.
\python and C++ have significant differences in the treatment of variables and memory:
While \python passes variables as a reference and does automatic garbage collection, C++ requires manual memory management by default.
With the introduction of C++11~\cite{ISO:2012:III} smart pointers, which provide memory management, have become available. Using smart pointers to store all 
the \EUDAQII objects instead of raw pointers bridges the gap between \python and C++. However, a small performance penalty has to be paid using \python due the use of an interpreter and an additional layer, which encapsulates the C++ methods and converts the interfaces to \python.
The \python interface enables a safe transfer of the \EVENT and \STATUS objects from C++ into \python or vice versa and a \python example
is provided as a part of the \EUDAQII package.

\section{User examples}
In most user cases, providing the dedicated and detector-specific user code and building them into \EUDAQII as a modular plugin is sufficient.
A modular plugin usually involves following parts:
\begin{enumerate}
\item Implementation of a set of \Producers for each new detector.
\item Implementation of a \DATACONVERTER to convert each \RAWDATAEVENT to a \STDDATAEVENT and to the \LCIO format, if used for later analysis.
\item Optionally, implementation of a  \DataCollector, if a specific synchronization, other than trigger-ID or timestamp based methods, is required.
\end{enumerate}
The \EUDAQII manual~\cite{Liu:2314266} provides detailed technical information as well as a set of code examples.

\EUDAQII contains many new features from long-standing requests of the user community that has used \EUDAQ for nearly a decade.
Several user groups have become early adopters already during the development phase of \EUDAQII and provided very valuable feedback throughout 
and are now using \EUDAQII for their test beam campaigns. 

Already today, \EUDAQII supports many test beam campaigns of detector prototypes at CERN and DESY.  
\EUDAQII is becoming more popular and many users migrate their \EUDAQ implementations to \EUDAQII .
Some examples for the usage of \EUDAQII are given below, showcasing the wide range of applications, ranging from the 
\EUDET-type and \LYCORIS beam telescopes to test beams for HL-LHC and CALICE. 

\subsection{\EUDET-type beam telescopes}
\label{sec:telescopes}

As the origin of the \EUDAQ framework, measurements based on \EUDET-type beam telescopes are the most extensive application of \EUDAQII /\EUDAQ software. 
Together with the \EUDET-TLU~\cite{tlu} they compose a common Trigger-DAQ infrastructure provided by several facilities.
Over the last decades, many different user setups have been integrated into \EUDAQ in order to use one of the seven copies of \EUDET-type beam 
telescopes located on five different beam lines \cite{eudaq12019}. 
In parallel, developments to achieve higher trigger rates in the Trigger-DAQ system of common beam telescopes 
\cite{behrens2015} have started.

\EUDET-type telescopes are \MIMOSAXXVI-based \cite{baudot2009, huguo2010}, which operate in a rolling shutter mode with a period of \SI{115.2}{\micro s}. 
A single FPGA receives the data of six individual telescope planes and creates trigger based double sensors frames. 
Trigger and rolling shutter are not correlated.
Thus, the next frame also needs to be read out resulting in a busy time of the telescope between \SI[parse-numbers=false]{115.2 \, and\, 230.4 }{\micro\second}.
A telescope event integrates all possible particle hits in the corresponding double frame read-out,
increasing the potential track multiplicity.
Using a faster pixel sensor, as for example a FE-I4~\cite{atlas-fei4} based sensor with a 25~ns read-out time,
allows identification of the track invoking the trigger by correlating the telescope track and the FE-I4 hit. 
Together with the global busy logic of the \EUDET-TLU, the maximum rate for tracks with a high time resolution is about 4~kHz. 

The development of \EUDAQII and the implementation of the \AIDA-TLU \cite{aidatlu_paper} allows for a new data taking mode which overcomes this trigger rate limit.
With the \EUDET-TLU, the system trigger rate has been limited by the slowest device.
This issue is overcome with the AIDA mode, that provides more flexibility by enabling individual configuration for each connected device.
By configuring an individual busy, faster devices can receive triggers while slower devices are still busy: 
A FE-I4 DAQ can potentially receive several trigger signal during the read-out time of the \MIMOSAXXVI-DAQ. 

The time information can be used to assign the multiple trigger information to potential multiple tracks in a \MIMOSAXXVI frame.
Applying this to the telescope system a trigger rate of 115~kHz is possible, or a factor of $\sim{28}$ of improvement, limited by the time required to read out the trigger ID. 
The trigger ID is used to synchronize the event streams in this data taking mode, by assigning unique numbers to triggers. 
The \AIDA-TLU provides the same data-trigger-busy communication protocol as the \EUDET-TLU to be compatible to existing device-under-test integration setups.

The \EUDAQ components of the telescope were upgraded and complemented within the \EUDAQII framework.
These are \Producers for the \MIMOSAXXVI DAQ and for both TLU types which define also the choice of the \DataCollector.
Operating with the \EUDET-TLU, the \texttt{EudetTluProducer} and the \texttt{EventIDSyncDataCollector} are used to synchronize 
the event streams by event number knowing that devices are operated in a trigger global busy scheme.
Operating with the \AIDA-TLU, the \texttt{AidaTluProducer} and the \texttt{TriggerIDSyncDataCollector} are used to synchronize event 
streams by trigger number which allows the new data taking mode as described above. 
Furthermore a generic \texttt{DirectSaveDataCollector} is provided which can be called multiple times, to store data from each connected  \Producer.
This is an example of distributed and decentralized data taking for which the event synchronization happens offline. 
Therefore, exemplary executables are provided for synchronizing multiple data streams by event number or trigger ID offline.
Finally, \DATACONVERTER modules can convert the TLU, FE-I4 and \MIMOSAXXVI \RAWDATAEVENT to \STDDATAEVENT blocks for providing interpretable data to the \texttt{StdEventMonitor}
which is the exported \EUDAQ \OnlineMonitor.  
Corresponding \DATACONVERTER modules are available to convert events to the \LCIO format~\cite{Aplin:2012kj} in order to perform 
track reconstruction with the \EUTELESCOPE framework \cite{eutel2019} as part of the \EUDET-type telescope infrastructure.

\subsection{ATLAS ITk Strip}

The ATLAS Inner Tracker (ITk) is a planned silicon tracker which is foreseen to start operation in 2026.
It comprises an inner section consisting of pixel sensors, and an outer section consisting of strip sensors.
The latter will cover an active area of approximately 165~m$^2$ with 17888 modules.
Each module is composed by a silicon sensor, front-end electronics and a power board.

Since 2017, five successful test beam campaigns with \EUDAQII have been conducted.
Two fully irradiated prototype end-cap module and the first double-sided prototype end-cap module were tested, among other devices \cite{Wiik}. 
The migration from \EUDAQ to \EUDAQII did not require extensive modifications to the previous setup.

As in \EUDAQ, two separate \Producers are used for the ATLAS ITk DAQ: a \Producer transmitting data from the front-end chips,
as the hit information, and a Producer providing TTC (Timing, Trigger and Control) information from the readout FPGA.
A dedicated converter to the \LCIO data format is implemented for each stream in order to perform track reconstruction
and data analysis using the \EUTELESCOPE reconstruction framework.

\subsection{\KPIX strip telescope}

The \KPIX readout system is used by the  \LYCORIS strip telescope~\cite{lycoris-D15.2} at the \diitbf.
Its native DAQ system consists of an ASIC called \KPIX~\cite{kpix}, and a FPGA DAQ board.
The \KPIX digitizes and serializes the collected data, the FPGA reads out the data from all the connected \KPIX and transmits to the PC.
The \KPIX chip is designed to be power cycled in between data acquisition periods, which requires an external acquisition start signal. 
The chip has an internal calibration and trigger module, but it can work with a forced external trigger. Both the 
acquisition start signal and the external trigger can be sent from the DAQ software.

\KPIX has its own DAQ software, which is the prototype version for the ROGUE~\cite{roguewebsite} framework developed by SLAC.
Given that \KPIX DAQ has been designed with multiple threads to control data taking and software command transmission,
the final integration to the \EUDAQII was implemented with an intermediate FIFO queue.
The \KPIX user module consists of
one \Producer connecting \KPIX via shared \KPIX DAQ libraries,
one \DataCollector linked with \KPIX binary data format libraries,
two \DATACONVERTER units to interpret a \RAWDATAEVENT to the \STDDATAEVENT and the \LCIO format,
and one \RunControl to ensure Stop and Reset functioning for all run modes with \KPIX.

Moreover, the \KPIX user module contains a customized GUI inherited from the core GUI for printing more detailed information like the Run Rate,
and one analysis executable called \texttt{lycorisCliConverter} for producing a set of straightforward plots like the ADC distribution of 
each channel to a \ROOT file from an \EUDAQII \RAWDATAEVENT.

\begin{figure}[htb]
  \begin{center}
    \includegraphics[width=0.65\textwidth]{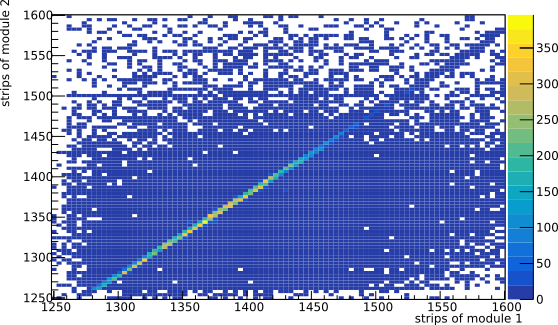}
    \caption{Spatial correlations of two \LYCORIS strip sensor modules from example data taken with the \EUDAQII~\KPIX user module at \diitbf. }
    \label{fig:kpix}
  \end{center}
\end{figure}

This user module has been used for data taking in multiple test beam campaigns. Figure~\ref{fig:kpix} shows the good spatial 
correlation between two \LYCORIS strip sensor modules taken with this \EUDAQII module.

\subsection{CALICE AHCAL}

The CALICE AHCAL is a prototype of a highly granular hadron calorimeter~\cite{Sefkow:2018rhp} optimized for the particle flow algorithm~\cite{pandorapfa} at a future 
\epem collider~\cite{ilc-tdr}. It uses $\SI[product-units = power]{30 x 30 x 3}{\milli\meter}$ scintillator tiles, which are individually 
read out by silicon-photomultiplier photodetectors.
The embedded, low-power readout electronics is based on the SPIROC chip~\cite{calice-conforti} with 36 input channels, specifically designed for a 
power-pulsing operation in sparse spills with an active duty cycle of less than $\SI{1}{\percent}$. 
The chip is self-triggered and stores for each channel a charge, a hit time and a bunch-crossing ID (BXID). 

The AHCAL \EUDAQII \Producer includes an configurable event building method,
producing events in various modes in order to be able to synchronize with other detectors:
\begin{description}
\item [{Timestamp}] is recorded in AHCAL DAQ hardware with $\SI{25}{\nano\second}$ resolution with the possibility to receive (or provide) 
the clock from (to) other systems. The timestamp is 48 bits wide and overflows only after 81 days.
\item [{Trigger number}] can be counted and timestamped from external source in the AHCAL DAQ hardware via a separate path. 
In the producer's event building stage, the trigger timestamp is then used to pair the trigger with the corresponding self-triggered 
hits from the ASIC data. The trigger ID is then used as an event number.
\item [{Acquisition cycle and BXID}] are unique identifiers, that can be used for synchronization with other acquisition-cycle oriented 
detectors, especially with front-end chips from the same family. 
The event can be build on a cycle level, containing all hits within the acquisition cycle (spill), or split into separate events by BXID.
\end{description}

\EUDAQII provides the possibility to run multiple data collectors. This feature was used
extensively during the synchronization studies with the \MIMOSA beam telescope, where several instances of a data collector with 
different event number offsets ran together. The event number offset was found by observing the spatial correlation in generated files. 
The events are merged in \EUDAQ by the event (trigger) number.

\begin{figure}[htb]
  \begin{center}
    \includegraphics[width=0.5\textwidth]{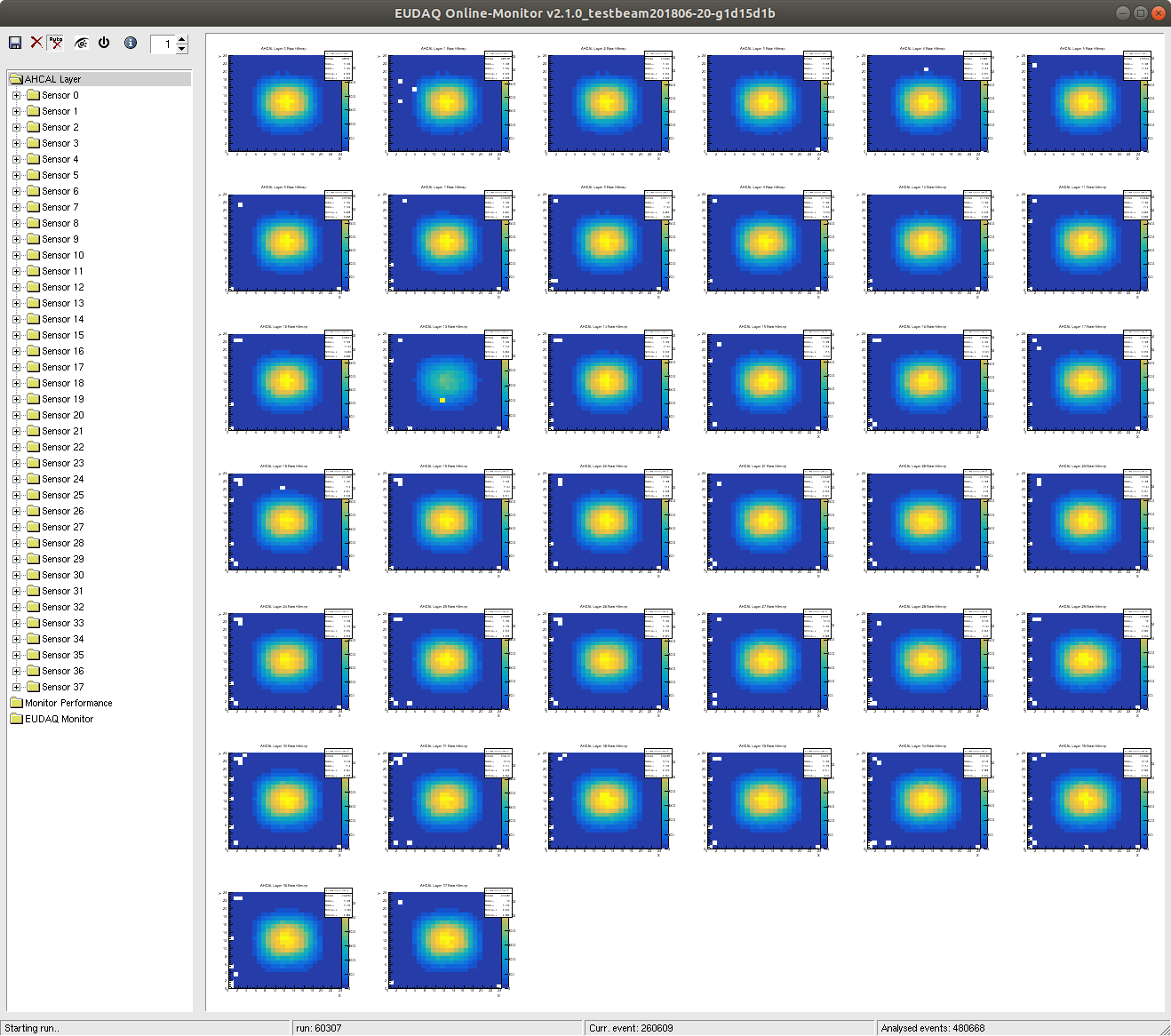}
    \caption{Hit maps of the AHCAL with 38 layers in the $\SI{40}{\giga\electronvolt} $ muon beam at the CERN SPS, May 2018.}
    \label{fig:ahcal_hitmap}
  \end{center}
\end{figure}

A data converter from AHCAL \EUDAQII  \RAWDATAEVENT to a \STDDATAEVENT provides a possibility to display the AHCAL plane ($\SI[product-units = power]{72x72}{cm}$) as 
a pixel plane in the \EUDAQII \OnlineMonitor, giving an immediate beam footprint feedback, as shown at Figure~\ref{fig:ahcal_hitmap}. 
The converter was used to check and monitor the spatial correlations with the \MIMOSA telescope at the \diitbf using the \OnlineMonitor.

The CALICE AHCAL has successfully used \EUDAQII in many commissioning and test beam campaigns. 
A specific AHCAL \RunControl automatically loading a new configuration before restart of a new run was used to perform internal calibration 
or automated position scans on the moving platform at the \diitbf.
Delay wire chambers were used at the CERN test beam for beam particle trajectory reference. 
Online event number synchronization (based on the trigger number) was not reliable and timestamps were not using the same time base. 
The synchronization was therefore achieved offline. 
The AHCAL used \EUDAQII together with the CMS-HGCAL~\cite{Collaboration:2293646} at the CERN SPS test beam in 2018. The event synchronization was done offline, 
based on the trigger numbers and validated with the timestamps.

\section{Conclusion}
\EUDAQII is a modular, modern and versatile data acquisition framework that has 
been developed as a successor of \EUDAQ. A core library is utilized to manage 
the readout, data collection and steering. The core only depends on standard C++ 
functionalities, allowing for platform independent developments. Hardware 
specific code, a \Producer, is linked to the core and can utilize more 
specific external libraries.

\EUDAQII is shipped together with \Producers for a trigger logic unit and the \EUDET-type telescopes, which are provided as common hardware on 
test beam facilities at DESY, ELSA and CERN.
The \EUDAQII software, together with a detailed operation manual is available online under the LGPLv3 open-source licence.
Already in the development phase,  \EUDAQII has been used successfully by several test beam groups.
\Producers to integrate pixel and strip 
sensors as well a highly granular calorimeter prototypes have been implemented. \EUDAQII is in a unique position to repeat the success story of the first \EUDAQ.

\acknowledgments
This project has received funding from the European Union's Horizon 2020 Research
and Innovation programme under Grant Agreement no. 654168.
This work was in part supported by the Science and Technologies Facilities Council, UK.
Measurements leading to these results have been performed at the Test Beam Facility at DESY Hamburg (Germany),
a member of the Helmholtz Association (HGF).

\bibliographystyle{elsarticle-num}
\bibliography{references}

\end{document}